\documentclass[journal]{IEEEtran}
\usepackage{graphicx}
\usepackage{float}
\usepackage{graphicx}
\usepackage{subfigure}
\usepackage{color}
\usepackage{amssymb}
\usepackage{amsmath}
\usepackage{algorithm}
\usepackage{algorithmic}
\usepackage{bm}
\usepackage{comment}
\usepackage[colorlinks]{hyperref}

\begin{document}
\title{  A Secure Beamforming Design: When Fluid Antenna Meets NOMA}

\author{Lifeng Mai, Junteng Yao, Jie Tang,  Tuo Wu, Kai-Kit Wong, \emph{Fellow, IEEE},\\
Hyundong Shin, \emph{Fellow, IEEE}, and Fumiyuki Adachi, \emph{Life Fellow}, \emph{IEEE}
\vspace{-9mm}

\thanks{L. Mai is with the Electric Power Research Institute, China Southern Power Grid, Guangzhou 510663, China. He is also with the Guangdong Provincial Key Laboratory of Power System Network Security, Guangzhou, Guangdong, China (E-mail: $\rm mailf@csg.cn$). J. Yao is with the Faculty of Electrical Engineering and Computer Science, Ningbo University, Ningbo 315211, China (E-mail:  $\rm yaojunteng@nbu.edu.cn$). J. Tang is with the School of Electronic and Information Engineering, South China University of Technology, Guangzhou 510006, China (E-mail:  $\rm eejtang@scut.edu.cn$). T. Wu is with the School of Electrical and Electronic Engineering, Nanyang Technological University, 639798, Singapore (E-mail: $\rm tuo.wu@ntu.edu.sg$). K. K. Wong is affiliated with the Department of Electronic and Electrical Engineering, University College London, Torrington Place, WC1E 7JE, United Kingdom and he is also affiliated with the Department of Electronic Engineering, Kyung Hee University, Yongin-si, Gyeonggi-do 17104, Korea (E-mail.: $\rm kai\text{-}kit.wong@ucl.ac.uk$). H. Shin is with the Department of Electronic Engineering, Kyung Hee University, Yongin-si, Gyeonggi-do 17104, Korea (E-mail:  $\rm hshin@khu.ac.kr$). F. Adachi is with the International Research Institute of Disaster Science (IRIDeS), Tohoku University, Sendai, Japan (E-mail: $\rm adachi@ecei.tohoku.ac.jp$). }  
%\thanks{J. Yao is with the Faculty of Electrical Engineering and Computer Science, Ningbo University, Ningbo 315211, China (E-mail:  $\rm yaojunteng@nbu.edu.cn$).}
%\thanks{J. Tang is with the School of Electronic and Information Engineering, South China University of Technology, Guangzhou 510006, China (E-mail:  $\rm eejtang@scut.edu.cn$).} 
%\thanks{T. Wu and C. Yuen are with the School of Electrical and Electronic Engineering, Nanyang Technological University, 639798, Singapore (E-mail: $\rm \{tuo.wu, chau.yuen\}@ntu.edu.sg$).}
%\thanks{K. K. Wong is affiliated with the Department of Electronic and Electrical Engineering, University College London, Torrington Place, WC1E 7JE, United Kingdom and he is also affiliated with the Department of Electronic Engineering, Kyung Hee University, Yongin-si, Gyeonggi-do 17104, Korea (E-mail.: $\rm kai\text{-}kit.wong@ucl.ac.uk$).}
%\thanks{F. Adachi is with the International Research Institute of Disaster Science (IRIDeS), Tohoku University, Sendai, Japan (E-mail: $\rm adachi@ecei.tohoku.ac.jp$).}
\thanks{(\textit{Corresponding author: Kai Kit Wong.})}
}% <-this %
%\vspace{-8mm}
\markboth{}%
{Mai \MakeLowercase{\textit{et al.}}: Secure Beamforming in FAS-Assisted NOMA Systems}

\maketitle

\begin{abstract}
This letter proposes a secure beamforming design for downlink non-orthogonal multiple access (NOMA) systems utilizing fluid antenna systems (FAS). We consider a setup where a base station (BS) with $M$ fluid antennas (FAs) communicates to a cell-center user (CU) and a cell-edge user (CEU), each with a FA. The CU is the intended recipient while the CEU is regarded as a potential eavesdropper. Our aim is to maximize the achievable secrecy rate by jointly optimizing the secure beamforming vectors and the positions of FAs. To tackle this, we adopt an alternating optimization (AO) algorithm that  optimizes secure beamforming and the positions of the FAs iteratively while keeping the other variables fixed. Numerical results illustrate that when FAs meet NOMA, the proposed scheme greatly enhances the secrecy rate compared to conventional multiple-input single-output (MISO) fixed antenna NOMA systems and other benchmark schemes.
\end{abstract}
	

\maketitle

\vspace{-2mm}
\section{Introduction}
\IEEEPARstart{A}{nticipation} for sixth-generation (6G) networks continues to grow and the challenge of supporting a massive number of connections has risen. This has led to the surge of non-orthogonal multiple access (NOMA) technologies \cite{YLiu17,9806417} which allows multiple users to share the same physical channel simultaneously. NOMA enhances spectral efficiency and user throughput, making it an exciting technology for achieving the high-density device connectivity required by 6G.

However, NOMA systems face challenges when users have similar channel conditions, resulting in nearly identical channel gains. This similarity makes signal differentiation difficult for successive interference cancellation (SIC) decoding to be effective and greatly degrades its advantage over orthogonal multiple access (OMA) systems. To address this matter, fluid antenna systems (FAS) have emerged as a robust solution \cite{WongKK21, new2024tutorial, MaW23, XLai24, JYao24}. By dynamically adjusting antenna properties to optimize spatial diversity, FAS can overcome interference, help user differentiation and preserve the efficiency of NOMA. 

Recent advancements have seen FAS-assisted NOMA systems investigated, with focuses on diverse applications including sum-rate maximization \cite{WKNew24}, short-packet communications \cite{JZheng24}, fairness consideration \cite{Yao24}, integrated sensing and communication  (ISAC) \cite{Ghadi24_1}, and wireless powered communication (WPC) \cite{Ghadi24_2}. Despite these developments, the intrinsic feature of the NOMA scheme, which empowers multiple users to simultaneously receive the same signal, introduces significant security risks within wireless networks \cite{Pakravan-2023}. This architecture could inadvertently facilitate eavesdropper access, posing substantial threats to communication integrity and privacy.
	
To mitigate these risks and ensure physical layer security, the adoption of secure beamforming is essential. Secure beamforming minimizes signal leakage and prevents unauthorized interception, thereby safeguarding communications against potential eavesdroppers. Despite its importance, the integration of secure beamforming within FAS-assisted NOMA systems remains unexplored. In this letter, we delve into secure beamforming for downlink FAS-assisted NOMA systems with the objective of maximizing the secrecy capacity. Our investigation involves jointly optimizing the secure beamforming vectors and the positions of the transmit fluid antennas (FAs) to enhance the overall security and performance.

%\vspace{-2mm}
\section{System Model and Problem Formulation}
We consider a downlink FAS-assisted secure multiple-input single-output (MISO) NOMA system that consists of a base station (BS) with $M$ FAs, a cell-center user (CU) and a cell-edge user (CEU). Both users are equipped with a fixed-position antenna (FPA). The FAs at the BS are connected to $M$ radio frequency (RF) chains and each FA can change its position within some defined ranges, $\mathcal{S}_t$. The position change may not require mechanically movable antennas but be realized using electronically controlled RF pixels \cite{Zhang-2024ojcom}. The location of the $n$-th FA of the BS is defined using a two-dimensional Cartesian coordinate model and expressed as $\mathbf{t}_{m}=[x_m,y_m]^T, n\in\mathcal{M}=\{1,\dots,M\}$. The collective locations of the BS's  FAs are represented as $\mathbf{{t}} = [\mathbf{t}_1, \mathbf{t}_2, \dots, \mathbf{t}_M] \in \mathbb{R}^{2 \times M}$.    Moreover, within the system, the CU is assumed to be an entrusted user but the CEU is regarded as a potential eavesdropper, which may eavesdrop the information signals to the CU \cite{ZhangY16}.

\vspace{-2mm}	
\subsection{Signal Model}
In NOMA, the transmit signal from the BS is given by
\begin{align}\label{q0}
\mathbf{w}=\mathbf{w}_1s_1+\mathbf{w}_2s_2,
\end{align}
in which $\mathbf{w}_1 \in \mathbb{C}^{M\times 1}$, $\mathbf{w}_2 \in \mathbb{C}^{M\times 1}$, $s_1$ and $s_2$ denote the beamforming vectors and  the signals to the CU and CEU, respectively. Besides, by  denoting  $p_{k,m}$ as the power of the $m$-th antenna for user $k$, we have $\mathbb{E}[\mathbf{w}_1\mathbf{w}_1^{\dagger}]={\rm diag}\{p_{1,1}, \dots, p_{1,M} \}$ and $\mathbb{E}[\mathbf{w}_2\mathbf{w}_2^{\dagger}]={\rm diag}\{p_{2,1}, \dots, p_{2,M} \}$ for the CU and the CEU, respectively. Consequently, the received signals at the CU and the CEU are, respectively, written as
\begin{align}
{z}_c&=\mathbf{h}^{\dagger}_c(\mathbf{t})\mathbf{w}+{n}_c,\label{q1}\\ 
{z}_e&=\mathbf{h}^{\dagger}_e(\mathbf{t})\mathbf{w}+{n}_e,\label{q2}
\end{align}
where $\mathbf{h}_c(\mathbf{t})\in \mathbb{C}^{M\times 1}$ and $\mathbf{h}_e(\mathbf{t})\in \mathbb{C}^{M\times 1}$ represent the channel vectors between the BS and the CU, and the CEU, respectively. The expressions of $\mathbf{h}_c$ and $\mathbf{h}_c$ will be formulated in the following subsection. Furthermore, ${n}_c\sim \mathcal{CN}(0,\sigma^2)$ and ${n}_e \sim \mathcal{CN}(0,\sigma^2)$ denote the complex additive white Gaussian noises at the CU and the CEU, respectively.

\vspace{-2mm}
\subsection{Channel Model}
We assume that the size of flexible position region for FA is significantly smaller than the distance between the transmitter and receiver so that the far-field model is valid. The angles of arrival (AoA) and angles of departure (AoD) are assumed to remain constant for each propagation path. The numbers of transmit paths of the BS-CU and the BS-CEU links are all assumed as $L_t$, while the number of the received paths of these two links are denoted by $L_c$ and $L_e$, respectively. For the transmit paths, the elevation and azimuth AoDs of the $p$-th path are denoted as $\theta_k^p \in [0,\pi]$ and $\phi_k^p \in [0,\pi], k\in \{c,e\}$. In the $p$-th transmit path of the BS-CU and the BS-CEU links, the propagation distance difference between the position of the $m$-th FA and origin $\mathbf{t}_0=(0,0)^T$ is represented as
\begin{align}\label{q3}
\rho_k^p(\mathbf{t}_m)&=x_m\sin \theta_k^p \cos \phi_k^p + y_m\cos \theta_k^p, \ k\in\{c, e\}. 
\end{align}
Accordingly, the phase difference between the position of the $n$-th transmit FA and origin $\mathbf{t}_0=(0,0)^T$ in the $p$-th transmit path is $2\pi \rho_k^p(\mathbf{t}_n)/\lambda$, where $\lambda$ is the signal wavelength. Hence, the transmit field response vector can be written as
\begin{align}\label{q4}
\mathbf{g}_k\left(\mathbf{t}\right)&=\left[ \exp\left(\frac{j 2\pi \rho_k^1(\mathbf{t})}{\lambda}\right), \dots, \exp\left(\frac{j 2\pi \rho_k^{L_t}(\mathbf{t})}{\lambda}\right) \right]^T,
\end{align}
where $k\in\{c, e\}$. The transmit response matrix of all $M$ transmit FAs can be denoted as $\mathbf{G}_k\left(\mathbf{t}\right)\triangleq\left[ \mathbf{g}_k\left(\mathbf{t}_1\right), \dots, \mathbf{g}_k\left(\mathbf{t}_{M}\right) \right] \in \mathbb{C}^{L_t \times M}, k\in\{c,e\}$. For the receive paths, since the CU and CEU  are both equipped with FPA, the central and cell-edge field response vectors can be expressed as $\mathbf{f}_c \in \mathbb{C}^{L_c \times 1}$ and $\mathbf{f}_e \in \mathbb{C}^{L_e \times 1}$, where the elements of $\mathbf{f}_c$ and $\mathbf{f}_e$ are both $1$.

The path response matrix from the origin $\mathbf{t}_0=(0,0)^T$ to the CU and the the CEU are defined as $\mathbf{\Sigma} \in \mathbb{C}^{L_t \times L_c}$ and $\mathbf{\Omega} \in \mathbb{C}^{L_t \times L_e}$. We assume the response coefficients $\mathbf{\Sigma}_{q,p}$ and $\mathbf{\Omega}_{j,p}$ are independently and identically distributed (i.i.d.), which is modeled as a Gaussian distributed random variable with zero mean and variances $n_c^2$ and $n_e^2$.  Therefore, we have the channel matrices  $\mathbf{h}_c=\mathbf{G}_c^{\dagger}(\mathbf{t}) \mathbf{\Sigma}\mathbf{f}_c$ and $\mathbf{h}_e= \mathbf{G}_e^{\dagger}(\mathbf{t})\mathbf{\Omega}\mathbf{f}_e$.

\subsection{Transmission Rate}
The transmission rate of $s_2$ at the CU and CEU is given by
\begin{align}\label{q14}
R_{k,2}&=\log_2 \left( 1 + \frac{ \mathbf{w}_2^{\dagger} \mathbf{h}_k(\mathbf{t})\mathbf{h}^{\dagger}_k(\mathbf{t})\mathbf{w}_2}{\sigma^2 +  \mathbf{w}_1^{\dagger} \mathbf{h}_k(\mathbf{t})\mathbf{h}^{\dagger}_k(\mathbf{t})\mathbf{w}_1} \right), k\in\{c, e\}. 
\end{align}
By applying NOMA, the user decodes and removes $s_2$ with SIC. After that, the transmission rate of decoding $s_1$ at the CU and the CEU can be written as
\begin{align}\label{q16}
R_{k,1}&=\log_2 \left( 1 + \frac{\mathbf{w}_1^{\dagger} \mathbf{h}_k(\mathbf{t})\mathbf{h}^{\dagger}_k(\mathbf{t})\mathbf{w}_1}{\sigma^2} \right), \ k\in\{c, e\}.
\end{align}

\subsection{Problem Formulation}
The achievable secrecy rate of the CU under weak secrecy conditions can be written as
\begin{align}\label{q18}
R_s=R_{c,1}-R_{e,1}.
\end{align}
Our objective is to maximize the secrecy rate of the CU subject to a rate requirement at the CEU, a transmit power constraint at the BS and the position constraint of each transmit FA. Therefore, the optimization problem of FAS-assisted secure MISO-NOMA systems can be formulated as
\begin{subequations}\label{op1}
\begin{align}\label{op1a}
\max_{\mathbf{w}_1, \mathbf{w}_2, \mathbf{t}} &~R_s  \\
			\label{op1b} \mbox{s.t.} &~R_{k,2} \geq r, \ k \in \{c, e\},\\
			\label{op1c} &~\mathbf{w}_1^{\dagger}\mathbf{w}_1+\mathbf{w}_2^{\dagger}\mathbf{w}_2\leq P_{\max},\\
			\label{op1d} &~\mathbf{t} \in S_t,\\
			\label{op1e} &~\|\mathbf{t}_m-\mathbf{t}_k\|_2\geq D, \ m,k=1,\dots, M, \ m\neq k,
\end{align}
\end{subequations}
in which $r$ represents the transmission rate requirements, $D$ is the minimum distance between the FAs to prevent mutual coupling, and $P_{\max}$ is the maximum transmit power.

\vspace{-2mm}
\section{Secrecy Rate Maximization}
Here, we adopt the alternating optimization (AO) approach by iteratively optimizing the secure beamforming solution and the position of each FA with the other variables being fixed.

Letting $L^{(c)}_r=2^{r_c}-1$ and $L^{(e)}_r=2^{r_e}-1$, \eqref{op1b} becomes
\begin{align}\label{b1}
\mathbf{w}_2^{\dagger} \mathbf{h}_k(\mathbf{t}) \mathbf{h}^{\dagger}_k(\mathbf{t})\mathbf{w}_2 \geq
L^{(k)}_r \left( \mathbf{w}_1^{\dagger} \mathbf{h}_k(\mathbf{t})\mathbf{h}^{\dagger}_k(\mathbf{t})\mathbf{w}_1 +\sigma^2 \right),
\end{align}
where $\ k \in \{ c, e \}$. Then, after some mathematical derivations of the optimization function, \eqref{op1} can be further rewritten as
\begin{equation}\label{op2}
\max_{\mathbf{w}_1, \mathbf{w}_2, \mathbf{t}} \frac{\sigma^2 + \mathbf{w}_1^{\dagger} \mathbf{h}_c(\mathbf{t})\mathbf{h}^{\dagger}_c(\mathbf{t})\mathbf{w}_1}{\sigma^2 + \mathbf{w}_1^{\dagger} \mathbf{h}_e(\mathbf{t})\mathbf{h}^{\dagger}_e(\mathbf{t})\mathbf{w}_1}~
\mbox{s.t.}~\eqref{op1c}\text{--}\eqref{op1e} ~\&~ \eqref{b1}.
\end{equation}
%Next, we apply the AO algorithm to solve the above optimization problem.

\vspace{-2mm}
\subsection{Majorization-Minimization (MM) for Secure Beamforming}
 Observing \eqref{op2} and given $\mathbf{t}$, let us further introduce the slack variable $\tau\geq0$, accordingly, \eqref{op2} is equivalent to
\begin{subequations}\label{op2p}
\begin{align}\label{op2pa}
\max_{\mathbf{w}_1, \mathbf{w}_2, \tau}&~\tau \\
			\label{op2pb} \mbox{s.t.} &~ \tau\le \frac{\sigma^2 + \mathbf{w}_1^{\dagger} \mathbf{h}_c(\mathbf{t})\mathbf{h}^{\dagger}_c(\mathbf{t})\mathbf{w}_1}{\sigma^2 + \mathbf{w}_1^{\dagger} \mathbf{h}_e(\mathbf{t})\mathbf{h}^{\dagger}_e(\mathbf{t})\mathbf{w}_1},\\
%			 (\sigma^2 &+ \mathbf{w}_1^{\dagger} \mathbf{h}_e(\mathbf{t})\mathbf{h}^{\dagger}_e(\mathbf{t})\mathbf{w}_1)  \leq \sigma^2 + \mathbf{w}_1^{\dagger} \mathbf{h}_c(\mathbf{t})\mathbf{h}^{\dagger}_c(\mathbf{t})\mathbf{w}_1,   \\
			\label{op2pc} &~ \eqref{op1c} ~\&~ \eqref{b1}.
\end{align}
\end{subequations}
By introducing the slack variable $\epsilon_k \geq 0$, the constraint \eqref{op2pb} can be equivalently rewritten as two constraints:
\begin{align}\label{bp1}
\sigma^2 + \mathbf{w}_1^{\dagger} \mathbf{h}_c(\mathbf{t})\mathbf{h}^{\dagger}_c(\mathbf{t})\mathbf{w}_1 &\geq \tau \epsilon_e, \\ \label{bp2}
\sigma^2 + \mathbf{w}_1^{\dagger} \mathbf{h}_e(\mathbf{t})\mathbf{h}^{\dagger}_e(\mathbf{t})\mathbf{w}_1 &\leq \epsilon_e.
\end{align}
The constraint \eqref{bp2} is convex but \eqref{bp1} is nonconvex because of  the convex term on the left side of the inequality, while \eqref{bp1} has the bilinear product $\tau \epsilon_e$ on the right side. We firstly solve the bilinear product, and have $\tau \epsilon_e \leq \psi(\tau, \epsilon_e; \tau^{(l)} \epsilon_e^{(l)}) = \frac{1}{4} ( \tau + \epsilon_e )^2 - \frac{1}{4} ( \tau^{(l)} - \epsilon_e^{(l)} )^2 - \frac{1}{2} ( \tau^{(l)} - \epsilon_e^{(l)} ) ( \tau - \tau^{(l)} - \epsilon_e + \epsilon_e^{(l)} )$, which is the first-order Taylor expansion of $- \frac{1}{4} ( \tau - \epsilon_e )^2$ at the $l$-th iteration point $( \tau^{(l)}, \epsilon_e^{(l)} )$. As for the convex term of \eqref{bp1}, we define a first-order Taylor expansion for $\mathbf{w}_1$ as
\begin{multline}
f_c(\mathbf{w}_1, \mathbf{w}_{1,(l)})=\\
2 \mathrm{Re}\{ \mathbf{w}_{1,(l)}^{\dagger} \mathbf{h}_c(\mathbf{t})\mathbf{h}^{\dagger}_c(\mathbf{t})\mathbf{w}_1 \} - \mathbf{w}_{1,(l)}^{\dagger} \mathbf{h}_c(\mathbf{t})\mathbf{h}^{\dagger}_c(\mathbf{t})\mathbf{w}_{1,(l)}.
\end{multline}
Thus, \eqref{bp1} in the $(l+1)$-th iteration is equivalent to
\begin{align}\label{bp6}
\psi(\tau, \epsilon_e; \tau^{(l)}, \epsilon_e^{(l)}) \leq \sigma^2 + f_c(\mathbf{w}_1, \mathbf{w}_{1,(l)}).
\end{align}
Similar to \eqref{bp6}, the constraint \eqref{b1} can be rewritten as
\begin{align}\label{bp7}
L^{(k)}_r \left( \mathbf{w}_1^{\dagger} \mathbf{h}_k(\mathbf{t})\mathbf{h}^{\dagger}_k(\mathbf{t})\mathbf{w}_1 +\sigma^2 \right) \leq f_k(\mathbf{w}_2, \mathbf{w}_{2,(l)}).%, \ k \in \{ c, e \}.
\end{align}
Therefore, given $\mathbf{w}_{1,(l)}$, $\mathbf{w}_{2,(l)}$, $\tau^{(l)}$ and $\epsilon_e^{(l)}$,  problem \eqref{op2p} in the $(l+1)$-th iteration can be recast into
\begin{equation}\label{op3p}
\max_{\mathbf{w}_1, \mathbf{w}_2, \tau, \epsilon_e} \tau~\mbox{s.t.}~\eqref{bp2}, \eqref{bp6}, \eqref{bp7} ~\&~ \eqref{op1c},
\end{equation}
which is convex and can be solved by CVX \cite{Boyd}.

\vspace{-2mm}
\subsection{Optimization of the $m$-th FA Positions}
Given $\mathbf{w}_1$, $\mathbf{w}_2$ and $\{ \mathbf{t}_n \}_{n\neq m}$, let us further introduce   slack variables $\tau\geq0$ and $\epsilon_k \geq 0$, the optimization problem can be accordingly reformulated as 
\begin{subequations}\label{op7}
	\begin{align}\label{op7a}
		\max_{\mathbf{t}_m, \tau, \epsilon_e}\ \ &\tau\  \\
		\label{op7b} \mbox{s.t.} \ \ \ &\eqref{bp1}, \eqref{bp2}, \eqref{op1b}, \eqref{op1d}~ \&~  \eqref{op1e}.
	\end{align}
\end{subequations}
Letting $\mathbf{V}_c= \mathbf{\Sigma}^{\dagger}\mathbf{f}_c \mathbf{f}_c^{\dagger} \mathbf{\Sigma} \in \mathbb{C}^{L_t \times L_t}$,$\mathbf{V}_e = \mathbf{\Omega}^{\dagger}\mathbf{f}_e \mathbf{f}_e^{\dagger} \mathbf{\Omega} \in \mathbb{C}^{L_t \times L_t}$ and $\mathbf{t}=\{\mathbf{t}_1, \dots, \mathbf{t}_M \}\in\mathbb{C}^{M \times 2}$, \eqref{q18} and \eqref{q14} can be written as 
\begin{align}\label{cp3}
	R_s&=\log_2 \left( \frac{\sigma^2+d_{c,1}(\mathbf{t})}{\sigma^2+d_{e,1}(\mathbf{t})} \right),\\ \label{cp4}
	R_{k,2}&=\log_2 \left( 1 + \frac{d_{k,2}(\mathbf{t})}{\sigma^2+d_{k,1}(\mathbf{t})} \right), k \in \{c, e\}
\end{align} 
where $d_{k,q}(\mathbf{t})=\mathbf{w}_q^{\dagger}\mathbf{G}_k^{\dagger} (\mathbf{t}) \mathbf{V}_k \mathbf{G}_k (\mathbf{t}) \mathbf{w}_q$. Further, we have
\begin{align}\label{cp5}
d_{k,q}(\mathbf{t})= \sum_{i=1}^M \sum_{j=1}^M w_{q,i}^{'} w_{q,j} \mathbf{g}_k^{\dagger}(\mathbf{t}_i) \mathbf{V}_k \mathbf{g}_k(\mathbf{t}_j).
\end{align}
Now, we can set $\xi_{i,j}^q=w_{q,i}^{'} w_{q,j}$ and $d_{k,q}(\mathbf{t}_m)=\sum_{n=1}^M \xi_{n,m} \mathbf{g}_k^{\dagger}(\mathbf{t}_n^{(l)}) \mathbf{V}_k \mathbf{g}_k(\mathbf{t}_m)$, by utilizing the Taylor expansion, the lower bound of $d_{k,q}(\mathbf{t}_m)$ in the $(l+1)$-th iteration can be given by $d_{k,q}(\mathbf{t}_m)\geq d_{k,q}^l(\mathbf{t}_m)=2\mathrm{Re}\{\sum_{n=1}^M \xi_{n,m} \mathbf{g}_k^{\dagger}(\mathbf{t}_n^{(l)}) \mathbf{V}_k \mathbf{g}_k(\mathbf{t}_m)\} + W_{k,q}^m$,
where $W_{k,q}^m= \sum_{i\neq m} \sum_{j\neq m} \xi_{i,j} \mathbf{g}_k^{\dagger}(\mathbf{t}_i) \mathbf{V}_k \mathbf{g}_k(\mathbf{t}_j) - \xi_{m,m} \mathbf{g}_k^{\dagger}(\mathbf{t}_m^{(l)}) \mathbf{V}_k \mathbf{g}_k(\mathbf{t}_m^{(l)})$. Then we define $\bm{\xi}_{m}^q=[\xi_{1,m}^q, \dots, \xi_{M,m}^q]$ and $\mathbf{Q}_{k,q}=\bm{\xi}_{m}^q \mathbf{G}_k^{\dagger}(\mathbf{t}^{(l)}) \mathbf{V}_k$. As such, the first term of lower bound can be expressed as $b_{k,q}(\mathbf{t}_m)=2 \mathrm{Re}\{\mathbf{Q}_k^q \mathbf{g}_k(\mathbf{t}_m)\}$. By utilizing the second-order Taylor expansion approach, the lower bound of $b_{k,q}(\mathbf{t}_m)$ can be transformed into
\begin{align}\label{cp10}
	b_{k,q}(\mathbf{t}_m) &\geq b_{k,q}^l(\mathbf{t}_m) = b_{k,q}(\mathbf{t}_m^{(l)}) + \nabla b_{k,q} (\mathbf{t}_m^{(l)}) \left( \mathbf{t}_m - \mathbf{t}_m^{(l)} \right) \nonumber \\
	&\quad - \frac{\beta_{k,m}}{2} \left( \mathbf{t}_m - \mathbf{t}_m^{(l)} \right)^T \left( \mathbf{t}_m - \mathbf{t}_m^{(l)} \right),
\end{align}
where $b_{k,q}(\mathbf{t}_m)=2 \mathrm{Re}\{\mathbf{Q}_k^q \mathbf{g}_k(\mathbf{t}_m)\}=\sum_{k=1}^{L_t} 2 |q_k| \cos (\zeta_k(\mathbf{t}_m))$, $\zeta_k(\mathbf{t}_m)=\frac{2\pi}{\lambda}\rho_k^p(\mathbf{t}_m)-\angle \mathbf{Q}_k^q$, $\beta_{k,m} =\frac{16 \pi^2}{\lambda^2}\sum_{k=1}^{L_t} |q_k|$, $ \nabla b_{k,q} (\mathbf{t}_m) = \left[ \frac{\partial b_{k,q} (\mathbf{t}_m) }{\partial x_m }, \frac{\partial b_k (\mathbf{t}_m) }{\partial y_m } \right]^T$, $\frac{\partial b_{k,q} (\mathbf{t}_m) }{\partial x_m }  =  - \frac{4\pi}{\lambda} \sum_{k=1}^{L_t} |q_k| \sin \theta_t^k \cos \phi_t^k  \sin (\zeta_k(\mathbf{t}_m))$, and $\frac{\partial b_{k,q} (\mathbf{t}_m) }{\partial y_m }  = - \frac{4\pi}{\lambda} \sum_{k=1}^{L_t} |q_k| \cos \theta_t^k \sin (\zeta_k(\mathbf{t}_m))$. Also, \eqref{cp10} is convex with respect to $\mathbf{t}_m$, so the constraint \eqref{bp1} in the $(l+1)$-th iteration is given by
\begin{align}\label{cp11}
	b_{c,1}^l(\mathbf{t}_m) + W_{c,1}^m + \sigma^2 \geq \psi(\tau, \epsilon_e; \tau^{(l)}, \epsilon_e^{(l)}).
\end{align}
Letting $\mu_{i,j}=w_{1,i}^{'} w_{1,j} - L_r w_{2,i}^{'} w_{2,j}$, \eqref{op1b} can be transformed into $\sum_{i=1}^M \sum_{j=1}^M \mu_{i,j} \mathbf{g}_k^{\dagger}(\mathbf{t}_i) \mathbf{V}_k \mathbf{g}_k(\mathbf{t}_j)\geq L_r \sigma^2$. Similar to \eqref{cp11}, we define $\bm{\mu}_{m}=[\mu_{1,m}^q, \dots, \mu_{M,m}^q]$, $\mathbf{B}_{k}=\bm{\mu}_{m} \mathbf{G}_k^{\dagger}(\mathbf{t}^{(l)}) \mathbf{V}_k$ and $z_{k}(\mathbf{t}_m)=2 \mathrm{Re}\{\mathbf{B}_k \mathbf{g}_k(\mathbf{t}_m)\}$, so the constraint \eqref{op1b} in the $(l+1)$-th iteration is equivalent to
\begin{align}\label{cp12}
	z_k^l(\mathbf{t}_m) + W_{k}^m \geq L_r \sigma^2, k \in \{c,e\},
\end{align}
where $z_k^l(\mathbf{t}_m)$ and $W_{k}^m$ are similar to \eqref{cp10}. As for \eqref{bp2}, we have the upper bound \cite{JSong15}
\begin{align}\label{cp13}
	&\mathbf{g}_e^{\dagger} (\mathbf{t}_m) \mathbf{V}_e \mathbf{g}_e (\mathbf{t}_m) \leq \mathbf{g}_e^{\dagger}(\mathbf{t}_m^{(l)}) (\mathbf{\Phi}_e-\mathbf{V}_e) \mathbf{g}_e(\mathbf{t}_m^{(l)}) \\ \nonumber
	&\qquad + \mathbf{g}_e^{\dagger} (\mathbf{t}_m) \mathbf{\Phi}_e \mathbf{g}_e (\mathbf{t}_m) - 2 \mathrm{Re}\{\mathbf{g}_e^{\dagger}(\mathbf{t}_m^{(l)}) (\mathbf{\Phi}_e-\mathbf{V}_e) \mathbf{g}_e(\mathbf{t}_m)\},
\end{align}
where $\mathbf{\Phi}_e=\lambda_{\max}^e \mathbf{I}_{L_t}$ and $\lambda_{\max}^e$ is the maximum eigenvalue of $\mathbf{V}_e$, respectively. From \eqref{cp13}, we know $\mathbf{g}_e^{\dagger} (\mathbf{t}_m) \mathbf{\Phi}_e \mathbf{g}_e (\mathbf{t}_m)=L_t \lambda_{\max}^e$, so the upper bound of $d_{e,1}(\mathbf{t}_m)$ is written as 
\begin{align}\label{cp14}
d_{e,1}(\mathbf{t}_m)&\leq d_{e,1}^u(\mathbf{t}_m)\notag\\
&=2\mathrm{Re}\{\mathbf{D}_{e,1} \mathbf{g}_k(\mathbf{t}_m)\} + 2 L_t\lambda_{\max}^e + W_{e,1}^m,
\end{align}
where $\mathbf{D}_{e,1}=\bm{\xi}_{m}^1 \mathbf{G}_e^{\dagger}(\mathbf{t}^{(l)}) \mathbf{V}_e-\xi_{m,m}^1 \mathbf{g}_e^{\dagger}(\mathbf{t}^{(l)}) \bm{\Phi}_e$. We define $c_{e,1}(\mathbf{t}_m)=2 \mathrm{Re}\{\mathbf{D}_{e,1} \mathbf{g}_k(\mathbf{t}_m)\}$. By utilizing the second-order Taylor expansion approach, the upper bound of $c_{e,1}(\mathbf{t}_m)$ can be transformed into
\begin{align}\label{cp15}
	c_{e,1}(\mathbf{t}_m)&\leq c_{e,1}^u(\mathbf{t}_m) = c_{e,1}(\mathbf{t}_m^{(l)}) + \nabla c_{e,1} (\mathbf{t}_m^{(l)}) \left( \mathbf{t}_m - \mathbf{t}_m^{(l)} \right) \nonumber \\
	&\qquad + \frac{\upsilon_{e,m}}{2} \left( \mathbf{t}_m - \mathbf{t}_m^{(l)} \right)^T \left( \mathbf{t}_m - \mathbf{t}_m^{(l)} \right),
\end{align}
where $c_{e,1}(\mathbf{t}_m)$, $\upsilon_{e,m}$ and $\nabla c_{e,1}(\mathbf{t}_m)$ are similar to $b_{k,q}(\mathbf{t}_m)$, $\beta_{k,m}$ and $\nabla b_{k,p}(\mathbf{t}_m)$. Therefore, the constraint \eqref{bp2} in the $(l+1)$-th iteration is equivalent to
\begin{align}\label{cp16}
	c_{e,1}^u(\mathbf{t}_m) + 2 L_t \lambda_{\max}^e + W_{e,1}^m+\sigma^2 \leq \epsilon_e.
\end{align}
Next we utilize the first-order Taylor expansion to relax $\|\mathbf{t}_m-\mathbf{t}_k\|_2$, and the constraint \eqref{op1e} in the $(l+1)$-th iteration can be expressed as
\begin{align}\label{cp17}
	\frac{1}{\| \mathbf{t}_m^{(l)} - \mathbf{t}_k \|_2} ( \mathbf{t}_m^{(l)} - \mathbf{t}_k )^T ( \mathbf{t}_m - \mathbf{t}_k ) \geq D.
\end{align}
Therefore, the optimization problem of $\mathbf{t}_m$ is rewritten as
\begin{subequations}\label{op8}
	\begin{align}\label{op8a}
		\max_{\mathbf{t}_m, \tau, \epsilon_e}\ \ &\tau\  \\
		\label{op8b} \mbox{s.t.} \ \ \ &\eqref{op1d}, \eqref{cp11}, \eqref{cp12}, \eqref{cp16}~ \& ~ \eqref{cp17}.
	\end{align}
\end{subequations}
Note that the above problem \eqref{op8} is convex with respect to $\mathbf{t}_m$, which can be solved using CVX \cite{Boyd}. We sequentially optimize each transmit FA position $\mathbf{t}_m$ until the algorithm converges to a fixed $\mathbf{t}$. Accordingly, we can obtain the locally optimal values $\mathbf{w}_1^{*}$, $\mathbf{w}_2^{*}$ and $\mathbf{t}^{*}$, respectively.

\section{Simulation Results}
\begin{figure}[t]
	\begin{minipage}{0.49\linewidth}
		\centering
		\includegraphics[width= \textwidth]{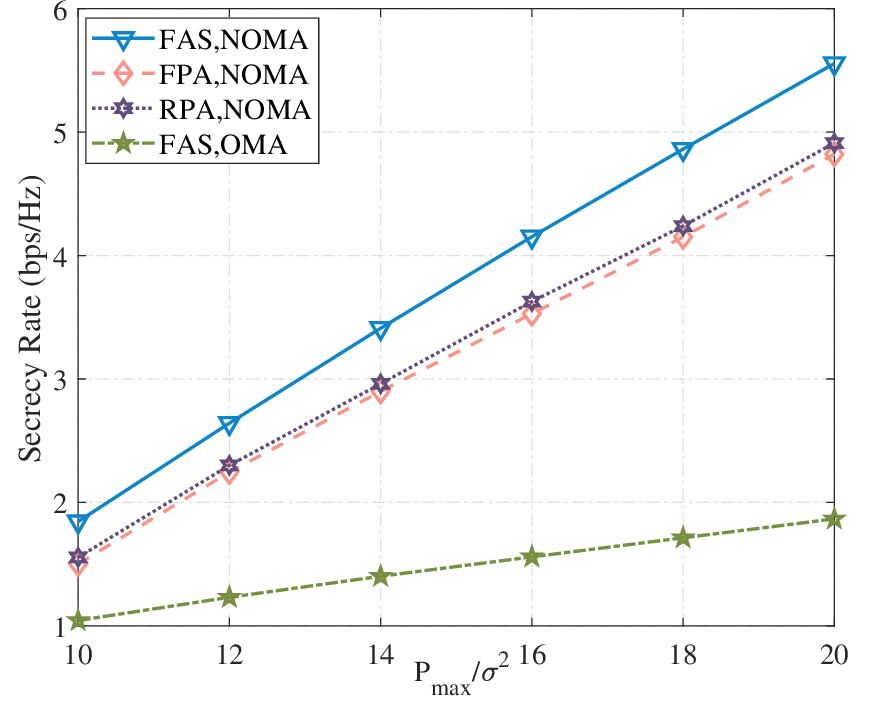}
		\caption{The secrecy rate versus $p_{\max}/\sigma^2$, where the constraint $r=2$ bps/Hz and $M=4$.}
		\label{fig2}
	\end{minipage}
	\hfill
	\begin{minipage}{0.50\linewidth} \vspace{-2mm}
		\centering
		\includegraphics[width= \textwidth]{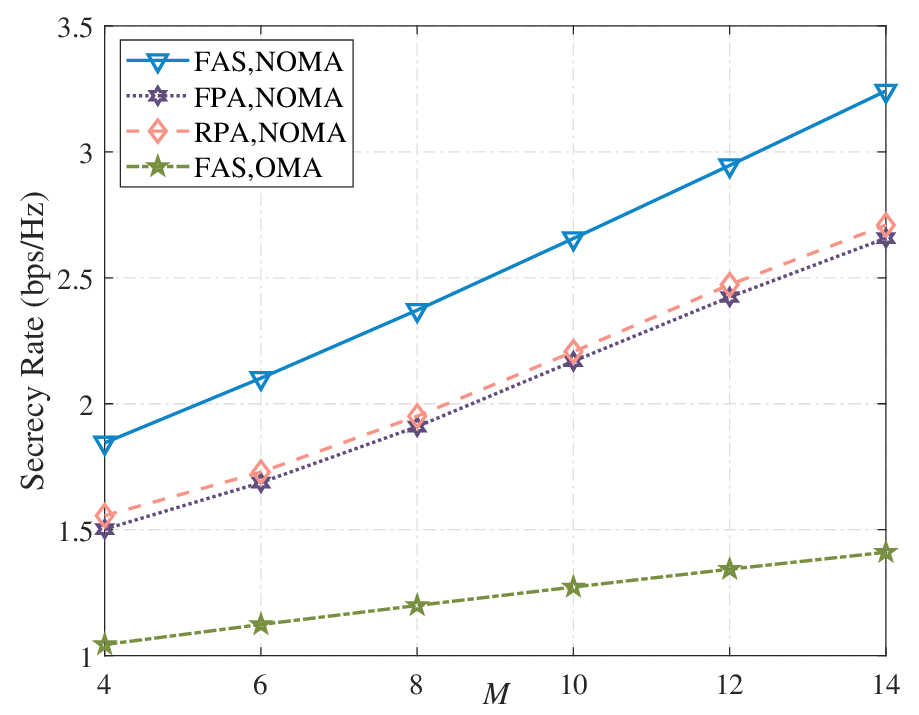}
		\caption{The secrecy rate versus antenna $M$, where the constraint $r=2$ bps/Hz and $p_{\max}/\sigma^2=10$ dB.} 
		\label{fig3}
	\end{minipage} 
\end{figure}

In the simulations, we set the number of receive antennas $N_c=N_e=1$. The distance between the CU/CEU and the BS is a uniform random variable, i.e., $d_k \sim [d_{\min}, d_{\max}], k \in \{c, e\}$, where $d_{\min}=20~{\rm m}$ and $d_{\max}=100~{\rm m}$ denote the nearest and the farthest distance, and $d_e \geq d_c$. We assume that the elevation and azimuth AoDs/AoAs of the BS $\{\theta_t^p\}_{p=1}^{L_t}$ and $\{\phi_t^p\}_{p=1}^{L_t}$ are i.i.d.~random variables in $[0, \pi]$, while that of users are fixed. The geometric channel model is adopted to describe the channel between the BS and the CU/CEU. The number of transmit and receive paths are identical, i.e., $L_t=L_c=L_e$, and we assume $L_t=4$. We define $\mathbf{\Sigma} \in \mathbb{C}^{L_t \times L_c}$ and $\mathbf{\Omega} \in \mathbb{C}^{L_t \times L_e}$, with the elements $\mathbf{\Sigma}_{q,p}\sim \mathcal{CN}(0, g_0 d_c^{-\alpha}/L)$ and $\mathbf{\Omega}_{j,p}\sim \mathcal{CN}(0, g_0 d_e^{-\alpha}/L))$, where $g_0 = -40~{\rm dB}$ is the average channel gain at the reference distance $d_0 = 1~{\rm m}$, $\alpha = 2.8$ is the path-loss exponent. The noise power is set as $\sigma_c^2=\sigma_e^2= -80~{\rm dBm}$. The carrier frequency is $2.4~{\rm GHz}$, which has a wavelength $\lambda=0.125~{\rm m}$. The minimum required distance between transmit/receive antennas is $D=\lambda/2$. Next we compare our proposed method with ``Random Position Antenna (RPA)" (i.e., randomly setting the antenna position in the given region while satisfying \eqref{op1e}) and ``FPA" (i.e., fixing the antenna position with the inter-spacing $D$).

Fig.~\ref{fig2} illustrates the secrecy rate performance of various methods across different values of $P_{\max}/\sigma^2$, where $r=2$ bps/Hz and $M=4$. As clearly shown, the proposed ``FAS, NOMA" scheme consistently demonstrates a superior secrecy rate compared to other approaches across different levels of $P_{\max}/\sigma^2$. Additionally, it is evident that the security rate achieved by ``FAS, NOMA" significantly surpasses that of ``FAS, OMA", highlighting the advantages of NOMA in terms of security enhancement within the FAS framework.

In Fig.~\ref{fig3}, we present the secrecy rate of different methods for different number of transmit antennas $M$, where we have the constraint $r=2~{\rm bps/Hz}$ and $p_{\max}/\sigma^2=10~{\rm dB}$. As can be observed, as $M$ increases from $4$ to $14$, the secrecy rate of our proposed ``FAS, NOMA" scheme outperforms others. It  highlights the benefits of NOMA and FAS for enhancing the security of wireless communication systems. This synergy between FAS and NOMA makes our scheme more robust against eavesdropping, as the higher secrecy rate suggests a stronger resilience to potential interception, offering valuable insights for secure communications in future networks.

\section{Conclusion}
In this letter, we studied the secure beamforming for FAS-assited NOMA systems, which was solved by iteratively optimizing secure beamforming and the position of each transmit FA. Numerical results have been provided to show that our proposed method is superior than others.

\end{document}